\documentstyle[12pt,epsf]{article}
\begin{document}\begin{flushright}\thispagestyle{empty}
OUT--4102--84\\
MZ--TH/99--57\\
hep-th/9912093\\
10 December 1999   \end{flushright}\vspace*{2mm}\begin{center}{
                                                    \Large\bf
Combinatoric explosion of renormalization tamed    \\[4pt]
by Hopf algebra: 30-loop Pad\'e-Borel resummation
                                                    }\vglue 10mm{\large{\bf
D.~J.~Broadhurst$^{1)}$ and D.~Kreimer$^{2)}$       }\vglue 4mm
Erwin Schr\"odinger Institute, A-1090 Wien, Austria }\end{center}\vfill
                                                  \noindent{\bf Abstract}\quad
It is easy to sum chain-free self-energy rainbows, to obtain contributions to
anomalous dimensions. It is also easy to resum rainbow-free self-energy chains.
Taming the combinatoric explosion of all possible nestings and chainings of a
primitive self-energy divergence is a much more demanding problem. We solve it
in terms of the coproduct $\Delta$, antipode $S$, and grading operator $Y$ of
the Hopf algebra of undecorated rooted trees. The vital operator is $S\star Y$,
with a star product effected by $\Delta$. We perform 30-loop Pad\'e-Borel
resummation of $463\,020\,146\,037\,416\,130\,934$ BPHZ subtractions in Yukawa
theory, at spacetime dimension $d=4$, and in a trivalent scalar theory, at
$d=6$, encountering residues of $S\star Y$ that involve primes with up to 60
digits. Even with a very large Yukawa coupling, $g=30$, the precision of
resummation is remarkable; a 31-loop calculation suggests that it is of order
$10^{-8}$.
\vfill\footnoterule\noindent
$^1$) D.Broadhurst@open.ac.uk;
http://physics.open.ac.uk/$\;\widetilde{}$dbroadhu\\
permanent address: Physics Dept, Open University, Milton Keynes MK7 6AA, UK\\
$^2$) Dirk.Kreimer@uni-mainz.de;
http://dipmza.physik.uni-mainz.de/$\;\widetilde{}$kreimer\\
Heisenberg Fellow, Physics Dept, Univ.\ Mainz, 55099 Mainz, Germany
\newcommand{\bookfig}[5]{\begin{figure}\centering\fbox{\epsfysize=#5cm
\epsfbox{#1}}\caption[#2]{\small #4}\label{#3}\end{figure}}
\newcommand{\ep}{\varepsilon}\newcommand{\de}{\delta}
\newpage\setcounter{page}{1}

\section{Introduction}

In this work we develop the Hopf algebra of
renormalization~\cite{DK,CK,Over,BK,Chen} to progress
beyond the rainbow~\cite{Del4,Del6} and chain~\cite{Chain,BG}
approximations for anomalous dimensions.

{\bf Summing rainbows:}\quad
In $d$ dimensions, the massless scalar one-loop integral
with propagators to the powers $\alpha,\beta$ is
\begin{equation}
G(\alpha,\beta;d):=g(\alpha)g(\beta)g(d-\alpha-\beta);\quad
g(\alpha):=\Gamma(d/2-\alpha)/\Gamma(\alpha)
\label{Gam}
\end{equation}
Now consider the interaction $g\phi^\dagger\sigma\phi$,
with a neutral scalar particle $\sigma$ coupled to a charged scalar $\phi$,
in the critical dimension, $d_c=6$.
To find the anomalous field dimension $\gamma$ of $\phi$, in the
rainbow approximation
of~\cite{Del6}, one solves the consistency condition
\begin{equation}
1=aG(1,1+\gamma;6)={a\over\gamma(\gamma-1)(\gamma-2)(\gamma-3)}
\label{cons}
\end{equation}
which ensures that the coupling $a:=g^2/(4\pi)^{d_c/2}$ cancels the
insertion of the anomalous self energy. The perturbative solution
of the resulting quartic is easily found:
\begin{equation}
\gamma_{\rm rainbow}=\frac{3-\sqrt{5+4\sqrt{1+a}}}{2}
=-\frac{a}{6}+11\frac{a^2}{6^3}-206\frac{a^3}{6^5}+\cdots
\label{R6}
\end{equation}

{\bf Resumming chains:}\quad
At the other extreme, one may
easily perform the Borel resummation of chains of self-energy
insertions, within a single rainbow. Suppose that the self energy
$p^2\overline\Sigma(a,p^2/\mu^2)$ is renormalized in the momentum scheme,
and hence vanishes at $p^2=\mu^2$. The renormalized massless propagator
is $\overline{D}=1/(p^2-p^2\overline\Sigma)$. Then~(\ref{R6}) is the
rainbow approximation for
$\partial\overline\Sigma/\partial\log(\mu^2)$ at
$p^2=\mu^2$. Following the methods of~\cite{Chain}, one finds that
the corresponding asymptotic series for chains is Borel resummable:
\begin{equation}
\gamma_{\rm chain}=-6\int_0^\infty{\exp(-6x/a)dx\over(x+1)(x+2)(x+3)}
\simeq-\frac{a}{6}+11\frac{a^2}{6^3}-170\frac{a^3}{6^5}+\cdots
\label{C6}
\end{equation}
which differs from the rainbow approximation at 3 loops, with
206 in~(\ref{R6}) coming from the triple rainbow,
while 170 in~(\ref{C6}) comes from a chain of two
self energies inside a third.

{\bf Hopf algebra:}\quad
We shall progress beyond the rainbow and chain approximations by including all
possible nestings and chainings of the one-loop self-energy divergence.
In other words, we consider the full Hopf algebra of undecorated rooted trees,
established in~\cite{DK} and implemented in~\cite{BK}.
Two figures suffice to exhibit the class of diagrams considered,
and their divergence structure. The first exhibits a 12-loop
example, the second exhibits its divergence structure.
Due to the fact that we combine chains and rainbows,
we have a full tree structure \cite{DK}: the depth of the tree is larger than
one, and there can be more than one edge attached to a vertex.
\bookfig{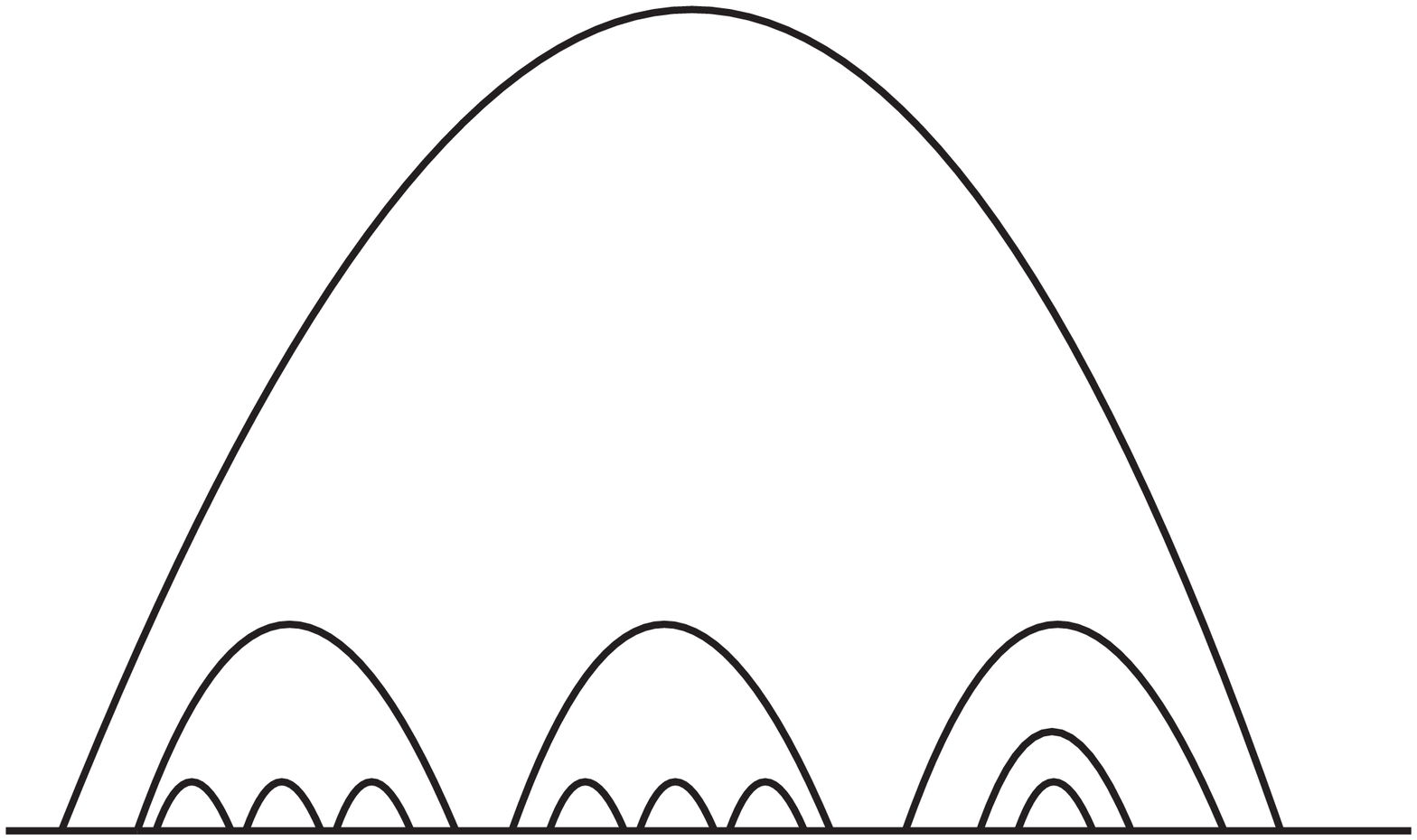}{12loops}{ddf3}{A 12-loop diagram based
on a one-loop skeleton.}{4}
\bookfig{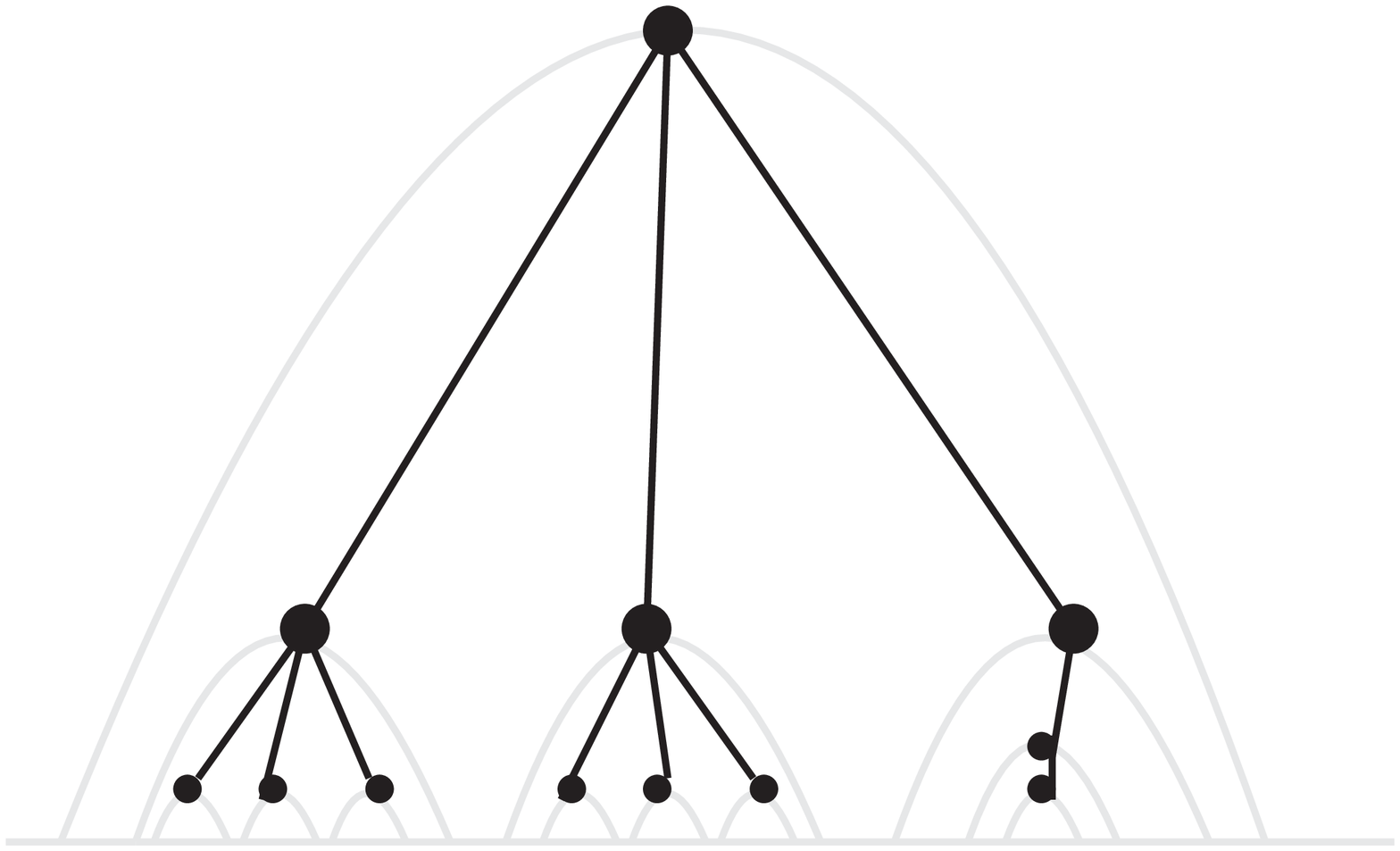}{6loops}{ddf4}{The divergence structure
of the previous figure.}{4}

There are 4 notable
features of this analysis.
\begin{enumerate}
\item We use the coproduct $\Delta$
to combine the antipode $S$ and grading operator
$Y$ in a star product $S\star Y$ whose residue
delivers the contribution of each rooted tree.
\item We show that the rationality of rainbows~\cite{Hab}
extends to the contribution of every undecorated rooted tree,
as had been inferred from examples in~\cite{KD}.
\item We confirm that a recent analysis~\cite{DR}
of dimensional regularization applies at both $d_c=4$ and $d_c=6$,
detecting poles of $\Gamma$ functions that occur
in even dimensions.
\item We obtain, to 30 loops, highly non-trivial alternating
asymptotic series, which we resum, to high precision,
by combining Pad\'e~\cite{BGK} and Borel~\cite{Chain,BG} methods.
\end{enumerate}

\section{Hopf-algebra method}

Let $t$ be an undecorated rooted tree,
denoting the divergence structure of
a Feynman diagram. Then its coproduct is defined, recursively, by
\begin{equation}
\Delta(t)=t\otimes e+id\otimes B_+(\Delta(B_-(t)))
\label{D}
\end{equation}
where $e$ is the empty tree, evaluating to unity,
$id$ is the identity map,
$B_-$ removes the root, giving a product of trees in general,
and $B_+$ is the inverse of $B_-$, combining products by restoring a
common root. The recursion terminates with
$\Delta(e)=e\otimes e\label{De}$
and develops a highly non-trivial structure
by the operation of the coproduct on products of trees
\begin{equation}
\Delta\left(\mbox{$\prod_k$}t_k\right)=\mbox{$\prod_k$}\Delta(t_k)
\label{prod}
\end{equation}
between each removal and restoration of a root.
In Sweedler notation, it takes the form
\begin{equation}
\Delta(t)=\mbox{$\sum_k$} a_k^{(1)}\otimes a_k^{(2)}
=t\otimes e+e\otimes t+
\mbox{$\sum_k^\prime$} a_k^{(1)}\otimes a_k^{(2)}
\label{sweedler}
\end{equation}
with single trees on the right and, in general,
products on the left. The
prime in the second summation indicates the absence of the empty tree.
The field-theoretic role of the coproduct is clear:
on the left products of subdivergences are identified;
on the right these shrink to points.
Subtractions are effected by the antipode,
defined by the recursion
\begin{equation}
S(t)=-t-\mbox{$\sum_k^\prime$} S(a_k^{(1)})a_k^{(2)}
\label{S}
\end{equation}
for a non-empty tree, with $S(\prod_k t_k)=\prod_k S(t_k)$
for products and $S(e)=e$.

Renormalization involves a twisted antipode, $S_R$.
Let $\phi$ denote the Feynman map that assigns a dimensionally regularized
bare value $\phi(t)$ to the diagram whose divergence
structure is labelled by the tree $t$.
Then we apply the recursive definition~\cite{BK}
\begin{equation}
S_R(t)=-R\left(\phi(t)+\mbox{$\sum_k^\prime$}S_R(a_k^{(1)})\phi(a_k^{(2)})
\right)
\label{SR}
\end{equation}
with a renormalization operator $R$ that sets $p^2=\mu^2$,
in both the momentum and MS schemes, and in the MS scheme
selects only the poles in $\ep:=(d_c-d)/2$.

We can use the coproduct to combine operators.
Suppose that $O_1$ and $O_2$ operate on trees and their products.
Then we define the star product $O_1\star O_2$ by
\begin{equation}
O_1\star O_2(t)=\mbox{$\sum_k$} O_1(a_k^{(1)})O_2(a_k^{(2)})
\label{star}
\end{equation}
with ordinary multiplication performed after $O_1$ operates
on the left and $O_2$ on the right of each term in the coproduct.
By construction, $S\star id$ annihilates everything
except the empty tree, $e$.
The presence of $R$ makes $S_R\star\phi$ finite
and non-trivial. In particular, the renormalized Green function is simply
\begin{equation}
\Gamma_R(t)=\lim_{\ep\to0}S_R\star\phi(t)
\label{GR}
\end{equation}
whose evaluation was efficiently encoded in~\cite{BK}, using
a few lines of computer algebra.

Here we present a new -- and vital --
formula for efficiently computing the contribution
of an undecorated tree to the anomalous dimension. It is simply
\begin{equation}
\gamma(t)=\lim_{\ep\to0} \ep \phi(S\star Y(t))
\label{ad}
\end{equation}
where $Y$ is the grading operator, with $Y(t)=n t$, for a tree with
$n$ nodes. In general, $Y$ multiplies a product of trees
by its total number of nodes.
To see that this works, consider the terms in~(\ref{GR}), in the momentum
scheme, before taking the limit $\ep\to0$. Each term has a momentum
dependence $\left(p^2\right)^{n(d-d_c)/2}$, where $n$ is the number of loops
(and hence nodes) of the tree on the right of the term in the Sweedler sum.
If we multiply by $n\ep$, and then let $\ep\to0$,
we clearly obtain the derivative w.r.t.\ $\log(\mu^2/p^2)$. Setting
$p^2=\mu^2$ we obtain the contribution to the anomalous dimension.
Thus $R$ plays no role and we may replace $S_R(t)$ by
$\lim_{R\to id}S_R(t)=\phi(S(t))$,
where $S$ is the canonical antipode. Multiplication by $n\ep$ is
achieved by $\ep\phi(Y(t))=n\ep\phi(t)$ on the right of the coproduct,
where $Y$ acts only on single trees. Hence the abstract operator
$S\star Y$ delivers the precise combination of products
of trees whose bare evaluation as Feynman diagrams is
guaranteed to have merely a $1/\ep$ singularity, with residue
equal to the contribution to the anomalous dimension.
Thus we entirely separate the combinatorics from the analysis.

\section{Example}

By way of example, we show how the 3-loop expansions of~(\ref{R6},\ref{C6})
result from~(\ref{ad}). The combinatorics are now clear.
The analysis, at first sight, seems to entail the detailed properties of
$\Gamma$ functions. However, appearances can be misleading.

In general, a dimensionally regularized bare value for a $n$-loop diagram,
corresponding to the undecorated rooted tree $t$,
is evaluated by the recursion~\cite{BK}
\begin{equation}
\phi(t)=\frac{L(\ep,n\ep)}{n\ep}\prod_k\phi(b_k)
\label{proc}
\end{equation}
where $b_k$ are the branches originating from the root of $t$. It terminates
with $\phi(e)=1$. For the scalar theory with $d_c=6$,
the master function is
\begin{equation}
L(\ep,\de)=
\frac{a\de}{\left(p^2\right)^\ep}
G(1,1+\de-\ep;6-2\ep)=
-\frac{a}{\left(p^2\right)^\ep}
{\Gamma(1-\de)\Gamma(1+\de)\Gamma(2-\ep)\over
\Gamma(4-\de-\ep)\Gamma(1+\de-\ep)}
\label{L6}
\end{equation}

Now the wonderful feature of~(\ref{ad}) is that it depends only on the
derivatives of $L(\ep,\de)$ w.r.t.\ $\de$ at $\ep=0$. This reflects
the fact that the anomalous dimension, unlike the Green
function, is insensitive to the details of the regularization method.
Thus we may, with huge savings in computation time,
replace the master function by
\begin{equation}
L(0,\de)=\frac{a}{(\de-1)(\de-2)(\de-3)}
=\sum_{n\ge0}g_n\de^n=-\frac{a}{6}+11\frac{a\de}{6^2}
-85\frac{a\de^2}{6^3}+O(\de^3)
\label{rat}
\end{equation}
which establishes that the contribution of each rooted tree is rational.
The residue of
the anomalous dimension operator $S\star Y$ feels only the
rational residues of $\Gamma$ functions; it is blind to the zeta-valued
derivatives that contribute to the renormalized Green function.

Now that the analysis has been drastically simplified, we return to the
combinatorics.
The double rainbow, $t_2$, has coproduct
$\Delta(t_2)=t_2\otimes e+e\otimes t_2+t_1\otimes t_1$
where $t_1$ is the single rainbow, with
$\Delta(t_1)=t_1\otimes e+e\otimes t_1$.
The antipodes are $S(t_1)=-t_1$ and $S(t_2)=-t_2+t_1^2$.
The star products are $S\star Y(t_1)=t_1$ and
$S\star Y(t_2)=2t_2-t_1^2$. Hence the contributions
to the anomalous dimensions are the residues of
$L(0,\ep)/\ep$ and $(L(0,2\ep)-L(0,\ep))L(0,\ep)/\ep^2$,
namely $g_0=-a/6$ and $g_1g_0=11a^2/6^3$.

Following this simple example, the reader should find it easy
to determine the anomalous dimension contributions of the two rooted trees
at 3 loops. For $t_3$, the triple rainbow graph, $S\star Y$
delivers $3t_3-3t_1t_2+t_1^3$, with residue
$g_2g_0^2+g_1^2g_0=-(85+11^2)a^3/6^5$, in agreement with~(\ref{R6}).
For the other diagram, $t_3^\prime$, with a double chain in a single
rainbow, it delivers
$3t_3^\prime-4t_1t_2+t_1^3$ with residue $2g_2g_0^2=-2\times85a^3/6^5$,
in agreement with~(\ref{C6}).
The Borel
resummation~(\ref{C6}) of chains corresponds to the result
$n!g_n g_0^n$ for a chain of $n$ self energies, inside a single rainbow.
Writing the anomalous dimension contribution
of the full Hopf algebra as the asymptotic series
\begin{equation}
\gamma_{\rm hopf}\simeq\sum_{n>0} G_n{(-a)^n\over6^{2n-1}}
\label{H6}
\end{equation}
we find that $G_3=3\times85+11^2=376$.

In this paper, we undertake Pad\'e-Borel resummation of
the full Hopf series~(\ref{H6}), to 30 loops. We also resum
\begin{equation}
\widetilde\gamma_{\rm hopf}\simeq\sum_{n>0}
\widetilde{G}_n{(-a)^n\over2^{2n-1}}
\label{H4}
\end{equation}
for the anomalous dimension of a fermion field
with a Yukawa interaction $g\overline\psi\sigma\psi$,
at $d_c=4$, whose rainbow approximation
\begin{equation}
\widetilde\gamma_{\rm rainbow}=1-\sqrt{1+a}
\label{R4}
\end{equation}
was obtained in~\cite{Del4}. At the other extreme, the
Borel-resummed chain approximation
\begin{equation}
\widetilde\gamma_{\rm chain}=-2\int_0^\infty{\exp(-2x/a)dx\over x+2}
\label{C4}
\end{equation}
is easily obtained from the
Yukawa generating function,
$\widetilde{L}(0,\de)=a/(\de-2)$.

\section{Results to 30 loops}

At 4 loops, there are 5 undecorated Wick contractions,
corresponding to 4 rooted trees,
one of which has weight 2. For the scalar theory, at $d_c=6$, the tally is
\begin{equation}
G_4=4890+4711+3595+3595+3450=20241=3^2\times13\times173
\label{G4}
\end{equation}
Already this becomes tedious to compute by hand.
Fortunately, the recursions~(\ref{D},\ref{S})
of the coproduct and antipode make it sublimely easy to automate
the procedure~(\ref{ad}).

At $n$ loops, the number of relevant Wick contractions
is the Catalan number $C_{n-1}$,
where $C_n:=\frac{1}{n+1}{2n\choose n}$.
At 30 loops, there are $C_{29}=1\,002\,242\,216\,651\,368$ contractions.
Symmetries reduce these to rooted
trees, with weights determined recursively by
$W(t)=w(t)\prod_k W(b_k)$ where $b_k$ are the branches obtained
by removing the root of $t$. The symmetry
factor of the root is $w(t)=(\sum_j n_j)!/\prod_j n_j!$
where $n_j$ is the number of branches of type $j$. The generating
formula for $R_n$, the number of rooted trees with $n$ nodes, is~\cite{EIS}
$\sum_{n>0}R_n x^n=x\prod_{n>0}(1-x^n)^{-R_n}$
which expresses the fact that removal of roots from all trees
with $n$ nodes
produces all products of trees with a total of $n-1$ nodes.
This gives $R_{30}=354\,426\,847\,597$. The number of terms
produced by applying the BPHZ procedure~\cite{BPHZ} to a single
tree with $n$ nodes is $2^{n}$.

{}From these enumerations, one finds -- with some trepidation --
that computation to 30 loops entails
$\sum_{n\leq30} 2^{n} R_n=463\,020\,146\,037\,416\,130\,934$
subtractions, each requiring 30 terms in its Laurent expansion,
with coefficients involving integers of $O(10^{60})$.
Brute force
would require processing of $O(10^{24})$ bits of data, which is
far beyond anything contemplated by current computer science.
The remedy is clear: recursion of coproduct and antipode,
to compute the residues of the anomalous dimension operator $S\star Y$.

Each new coproduct
or antipode refers to others with fewer loops. By storing these
we easily progressed to 13 loops, extending the sequence $G_n$ to
\begin{center}
$1,~11,~376,~20241,~1427156,~121639250,~12007003824,~1337583507153,~$\\
$165328009728652,~22404009743110566,~3299256277254713760,~$\\
$524366465815117346250,~89448728780073829991976$
\end{center}
For $\widetilde{G}_n$, in the Yukawa case, we obtained the 13-loop sequence
\begin{center}
$1,~1,~4,~27,~248,~2830,~38232,~593859,~10401712,~202601898$\\
$4342263000,~101551822350,~2573779506192$
\end{center}
At this point, recursion of individual trees hit a ceiling imposed by
memory limitations.

Beyond 13 loops, we stored
only the unique combination of terms that is needed
at higher loops, namely the momentum-scheme renormalized self energy.
Allocating 750~megabytes of main memory to Reduce~3.7~\cite{Red},
the time to reach 30 loops was 8 hours.
Of these, more than 2 hours were spent on garbage collection, indicating
the combinatoric complexity.
Results for the scalar and Yukawa theories are in Tables~1 and~2.
They are highly non-trivial. Factorization of
$G_{27}=2^6\times5\times103\times184892457645048836717
\times69943104850621681268329469624581$
needed significant use of Richard Crandall's elliptic curve
routine~\cite{REC}, while $G_{29}/240$
is a 60-digit integer that is most probably prime.

\section{Pad\'e-Borel resummation}

We combine Pad\'e-approximant~\cite{BGK}
and Borel-transformation~\cite{Chain,BG} methods. From~(\ref{C6})
we obtain the pure chain contribution
$G^{\rm chain}_{n+1}=(2^{n}+(2^{n}-1)3^{n+1})n!$ with, for example,
$G_4^{\rm chain}=(8+7\times81)\times6=3450$ appearing in~(\ref{G4})
as the smallest contribution of the 5 Wick contractions at 4 loops,
while the pure rainbow contribution, $4711$,
is next to largest. This is far removed from the situation at large $n$,
where the pure rainbow term is factorially smaller
than the pure chain term. At large $n$, we combine $C_{n-1}
\approx4^{n-1}/\sqrt{n^3\pi}$ Wick contractions, some of which
are of order $G^{\rm chain}_n$, while some are far smaller.
It is thus difficult to anticipate the large-$n$ behaviour
of $G_n$. We adopted an empirical approach,
finding that $S_n:=12^{1-n}G_n/\Gamma(n+2)$ varies little for $n\in[14,30]$,
as shown in the final column of Table~1. In the Yukawa case of Table~2,
we found little variation in $\widetilde{S}_n:=
2^{1-n}\widetilde{G}_n/\Gamma(n+1/2)$.

In the scalar case, at $d_c=6$,
Pad\'e-Borel resummation may be achieved by the Ansatz
\begin{equation}
\gamma_{\rm hopf}\approx-\frac{a}{12}\int_0^\infty P(ax/3)e^{-x}x^2\,dx
\label{P6}
\end{equation}
where $P(y)=1+O(y)$ is a $[M\backslash N]$ Pad\'e approximant,
with numerator $1+\sum_{m=1}^M c_m y^m$ and denominator
$1+\sum_{n=1}^N d_n y^n$, chosen so as to reproduce
the first $M+N+1$ terms in the asymptotic series~(\ref{H6}).
We expect $P(y)$ to have singularities only in the left half-plane.
In particular, a pole near $y=-1$ is expected, corresponding
to the approximate constancy of $S_n$ in Table~1. We fitted
the first 29 values of $G_n$ with a $[14\backslash14]$ Pad\'e
approximant $P(y)$, finding a pole at
$y\approx-0.994$. The other 13 poles have $\Re y<-1$. Moreover
there is no zero with $\Re y>0$. The test-value $G_{30}$ is reproduced to
a precision of $5\times10^{-16}$.

In the Yukawa case, at $d_c=4$, we made the Ansatz
\begin{equation}
\widetilde\gamma_{\rm hopf}\approx-\frac{a}{\sqrt{\pi}}\int_0^\infty Q(ax/2)
e^{-x}x^{1/2}\,dx;\quad Q(y):=\frac{\widetilde{P}(y)}{1+y}
\label{P4}
\end{equation}
suggested by Table~2. Here we put in by hand the suspected pole at $y=-1$.
The $[14\backslash14]$ approximant to $\widetilde{P}(y)=1+O(y)$
then has all its 14 poles at $\Re y<-1$ and no zero with
$\Re y>0$. The test-value $\widetilde{G}_{30}$ is reproduced to a
reassuring precision of $4\times10^{-17}$.

Table~3 compares resummation of the
full Hopf results~(\ref{H6},\ref{H4}) with
those from the far more restrictive
chain and rainbow subsets.
To test the precision of resummations~(\ref{P6},\ref{P4}),
we used the star product~(\ref{ad})
to perform the $2.6\times10^{21}$ BPHZ subtractions that yield the
exact 31-loop coefficients
\begin{eqnarray}
G_{31}&=&2^6\times3^3\times5\times139\times2957\times22279
\times69318820356301\times9602299922477621\nonumber\\&&{}
\times144927172127490232568467
\label{S31}\\
\widetilde{G}_{31}&=&2^5\times3^4\times5\times71\times109\times13224049649
\times473202021103152647613521
\label{Y31}
\end{eqnarray}
No change in the final digits of Table~3 results from using these.
At the prodigious Yukawa coupling $g=30$, corresponding
to $a=(30/4\pi)^2\approx5.7$, a $[15\backslash 15]$ Pad\'e approximant
gives $\widetilde\gamma_{\rm hopf}\approx-1.85202761$, differing by less
than 1 part in $10^8$ from the $[14\backslash14]$ result
$\widetilde\gamma_{\rm hopf}\approx-1.85202762$.
It appears that resummation
of undecorated rooted trees is under very good control,
notwithstanding the combinatoric explosion apparent
in~(\ref{S31},\ref{Y31}).

\section{Conclusions}

As stated in the introduction, we achieved 4 goals.
First, we found the Hopf-algebra construct~(\ref{ad})
that delivers undecorated contributions to anomalous dimensions.
Then we found that these are rational, with the $\Gamma$ functions
of~(\ref{L6}) contributing only their residues, via~(\ref{rat}).
Next, we exemplified the analysis of dimensional regularization
in~\cite{DR}, at two different critical dimensions, $d_c=6$
and $d_c=4$. The residues of a common set~(\ref{Gam}) of $\Gamma$ functions
determine both results. Finally, we obtained highly non-trivial
results, from all combinations of rainbows and chains, to 30 loops.
{\sl A priori}, we had no idea how these would compare with the
easily determined pure chain contributions. Tables~1 and~2 suggest
that at large $n$ the full Hopf-algebra results  exceed pure chains by
factors that scale like $n^22^n$ and $n^{1/2}2^n$, respectively.
Pad\'e approximation gave 15-digit agreement with exact 30-loop results.
In Table~3, we compare the Borel resummations~(\ref{P6},\ref{P4}) of the full
Hopf algebra with the vastly simpler rainbow
approximations~(\ref{R6},\ref{R4}) and the still rather trivial
chain approximations~(\ref{C6},\ref{C4}). Even at the very
large Yukawa coupling $g=30$ we claim 8-digit precision.
Apart from large-$N_f$ approximations~\cite{BGK}, we know of no other
large-coupling analysis of anomalous dimension contributions,
at spacetime dimensions $d\ge4$,
that progresses beyond pure rainbows~\cite{Del4,Del6}
or pure chains~\cite{Chain,BG}.

In conclusion: Hopf algebra tames the combinatorics of
renormalization, by disentangling the iterative subtraction of
primitive subdivergences from the analytical challenge of
evaluating dimensionally regularized bare values for Feynman
diagrams. Progress with the analytic challenge shall require the
expansion of skeleton graphs in the regularization parameter
$D-4$. After that, the Hopf algebra of decorated rooted trees
provides the tool to take care of the combinatorial challenge of
renormalization in general. Generalizations of the methods here to
cases where decorations are different, but still analytically
trivial, are conceivable. The results in \cite{shuffle} are of
this form. In the present case, where the combinatoric explosion
is ferocious, while the analysis is routine, the automation of
renormalization by Hopf algebra is a joy. How else might one resum
$2.6\times10^{21}$ BPHZ subtractions at 31 loops and achieve
8-digit precision at very strong coupling?

\noindent{\bf Acknowledgements:}\quad
This work was undertaken during the workshop {\sl Number Theory and Physics}
at the ESI in November 1999, where we enjoyed discussions with Pierre Cartier,
Werner Nahm, Ivan Todorov and Jean-Bernard Zuber. Past work with Alain Connes,
Bob Delbourgo, John Gracey and Andrey Grozin supports the present paper.

\raggedright

\vfill

\begin{center}{\bf Table~1:}\quad
Scalar coefficients in~(\ref{H6}),
with $S_n:=12^{1-n}G_n/\Gamma(n+2)$
\end{center}
$$\begin{array}{l|l|l}
n&G_n&S_n\\\hline
14&16301356287284530869810308&0.1165\\
15&3161258841758986060906197536&0.1177\\
16&650090787950164885954804021185&0.1186\\
17&141326399508139818539694443090940&0.1194\\
18&32389192708918228594003667471390750&0.1200\\
19&7805642594117634874205145727265669184&0.1205\\
20&1973552096478862083584247237907087008846&0.1209\\
21&522399387732959889862436502331522596697560&0.1212\\
22&144486332652501966354908665093390779463113660&0.1215\\
23&41681362292986022786933211385817840822702468640&0.1217\\
24&12520661507532542738174037622803485508817145773050&0.1218\\
25&3910338928202486568787314743084879349561179264255736&0.1220\\
26&1267891158800355844456289086726128521948839015617187260&0.1221\\
27&426237156086127437403654947366849019736474802601497417920&0.1221\\
28&148382376919675149120919349602375065827367635238832722748020&0.1222\\
29&53428133467243180546330391126922442419952183999220340144106320&0.1222\\
30&19876558632009586773182109989526780486481329823560105761256963720\!&0.1222
\end{array}$$

\newpage

\begin{center}{\bf Table~2:}\quad
Yukawa coefficients in~(\ref{H4}),
with $\widetilde{S}_n:=2^{1-n}\widetilde{G}_n/\Gamma(n+1/2)$
\end{center}
$$\begin{array}{l|l|l}
n&\widetilde{G}_n&\widetilde{S}_n\\\hline
14&70282204726396&0.3715\\
15&2057490936366320&0.3750\\
16&64291032462761955&0.3780\\
17&2136017303903513184&0.3806\\
18&75197869250518812754&0.3828\\
19&2796475872605709079512&0.3848\\
20&109549714522464120960474&0.3865\\
21&4509302910783496963256400&0.3880\\
22&194584224274515194731540740&0.3894\\
23&8784041120771057847338352720&0.3906\\
24&414032133398397494698579333710&0.3917\\
25&20340342746544244143487152873888&0.3928\\
26&1039819967521866936447997028508900&0.3937\\
27&55230362672853506023203822058592752&0.3946\\
28&3043750896574866226650924152479935036&0.3953\\
29&173814476864493583374050720641310171808&0.3961\\
30&10272611586206353744425870217572111879288\!&0.3968
\end{array}$$

\vfill

\begin{center}{\bf Table~3:}\quad
Comparison of chain, rainbow and full Hopf contributions
\end{center}
$$\begin{array}{l|lll|lll}
a&-\gamma_{\rm chain}&-\gamma_{\rm rainbow}&-\gamma_{\rm hopf}&
-\widetilde\gamma_{\rm chain}&
-\widetilde\gamma_{\rm rainbow}&
-\widetilde\gamma_{\rm hopf}\\\hline
0.5 & 0.0727579 & 0.0731322 & 0.0742476 & 0.2245593 & 0.2247449 & 0.2278233\\
1.0 & 0.1301409 & 0.1322419 & 0.1373080 & 0.4126913 & 0.4142136 & 0.4281423\\
1.5 & 0.1773375 & 0.1825988 & 0.1937609 & 0.5765641 & 0.5811388 & 0.6118625\\
2.0 & 0.2172313 & 0.2268615 & 0.2455916 & 0.7226572 & 0.7320508 & 0.7837372\\
2.5 & 0.2516214 & 0.2665867 & 0.2939133 & 0.8549759 & 0.8708287 & 0.9464649\\
3.0 & 0.2817148 & 0.3027756 & 0.3394353 & 0.9762193 & 1.0000000 & 1.1017856\\
3.5 & 0.3083635 & 0.3361156 & 0.3826462 & 1.0883141 & 1.1213203 & 1.2509126\\
4.0 & 0.3321923 & 0.3671015 & 0.4239016 & 1.1926947 & 1.2360680 & 1.3947383\\
4.5 & 0.3536734 & 0.3961033 & 0.4634712 & 1.2904639 & 1.3452079 & 1.5339452\\
5.0 & 0.3731724 & 0.4234058 & 0.5015652 & 1.3824908 & 1.4494897 & 1.6690711\\
5.5 & 0.3909778 & 0.4492331 & 0.5383523 & 1.4694751 & 1.5495098 & 1.8005504\\
6.0 & 0.4073216 & 0.4737658 & 0.5739698 & 1.5519895 & 1.6457513 & 1.9287404
\end{array}$$

\end{document}